\documentclass[dvips]{article}

\usepackage{icrc2011}

\title{The Disp Method for Analysing Large Zenith Angle Gamma-Ray Data}

\newcommand{\etal}{\MakeLowercase{\textit{et al. }}} 
\shorttitle{\c{S}ent\"{u}rk \etal The Disp Method}

\authors{G\"{u}ne\c{s} D. \c{S}ent\"{u}rk$^{1}$ for the VERITAS Collaboration$^{2}$ }
\afiliations{$^1$Physics Department, Columbia University, New York, NY 10027, USA\\ $^2$see J. Holder et al. (these proceedings) or http://veritas.sao.arizona.edu/conferences/authors?icrc2011 }
\email{gunessenturk@gmail.com}

\abstract{The Disp method is an algorithm that is used for reconstruction of primary gamma ray direction in ground-based atmospheric 
Cherenkov telescope experiments -measuring very-high-energy (VHE) gamma rays in the energy range between 100GeV and ~30 TeV. 
In general terms, the geometric information obtained from one single shower image is sufficient for the algorithm to find the sky location of the primary.
Various versions of the Disp method were implemented and used in the past.
In this study, we present a multi-dimensional implementation of the Disp method for the VERITAS instrument and show (using Monte Carlo simulations and the Crab Nebula observations) that it significantly improves the angular resolution for large-zenith-angle (LZA) observations. We also applied the disp method to VERITAS data taken from the galactic center region which is detected by VERITAS. }

\begin{document}
\maketitle

\section{Motivation}

The Disp method has been the default direction reconstruction algorithm for single telescope observations (e.g.~\cite{lessard01,kranich03,domingo05}). 
The introduction of new generation ground-based gamma-ray instruments operating in array mode with multiple telescopes allowed for new and more accurate techniques for direction reconstruction. 
However, due to a larger (on average) impact parameter of the air showers with respect to the center of the of the IACT array and projection effects, these (geometrical) reconstruction techniques do not perform well at LZA. 
The Disp method compensates for this loss in quality of the angular reconstruction.


A description of the VERITAS analysis steps prior to direction reconstruction (calibration, image cleaning and image parametrization) can be found in ~\cite{cogan08}.
At the latter step, a second moment parametrization with $Hillas$ parameters is performed on each shower image~\cite{hillas85}. 
See Figure 1b for a detailed representation of relevant $Hillas$ parameters.
\emph{Disp} is the angular distance between the image centroid and the real source location on the camera plane. 
Another important image parameter is \emph{size}, defined as the total integrated charge in image pixels, representing a measure of brightness for the image. 
Figure 1a shows the stereoscopic image of an air shower after parametrization. 
The default VERITAS algorithm looks for the point with the smallest total distance from each major axis: this is the reconstructed direction. 
The error associated with this calculation is inversely related to the angles between the major axes. 
In other words, images with close to parallel major axes (Figure 1c), which more often occur for LZA data, result in large errors in direction reconstrucion. 

 \begin{figure*}[th]
  \centering
  \includegraphics[width=2in]{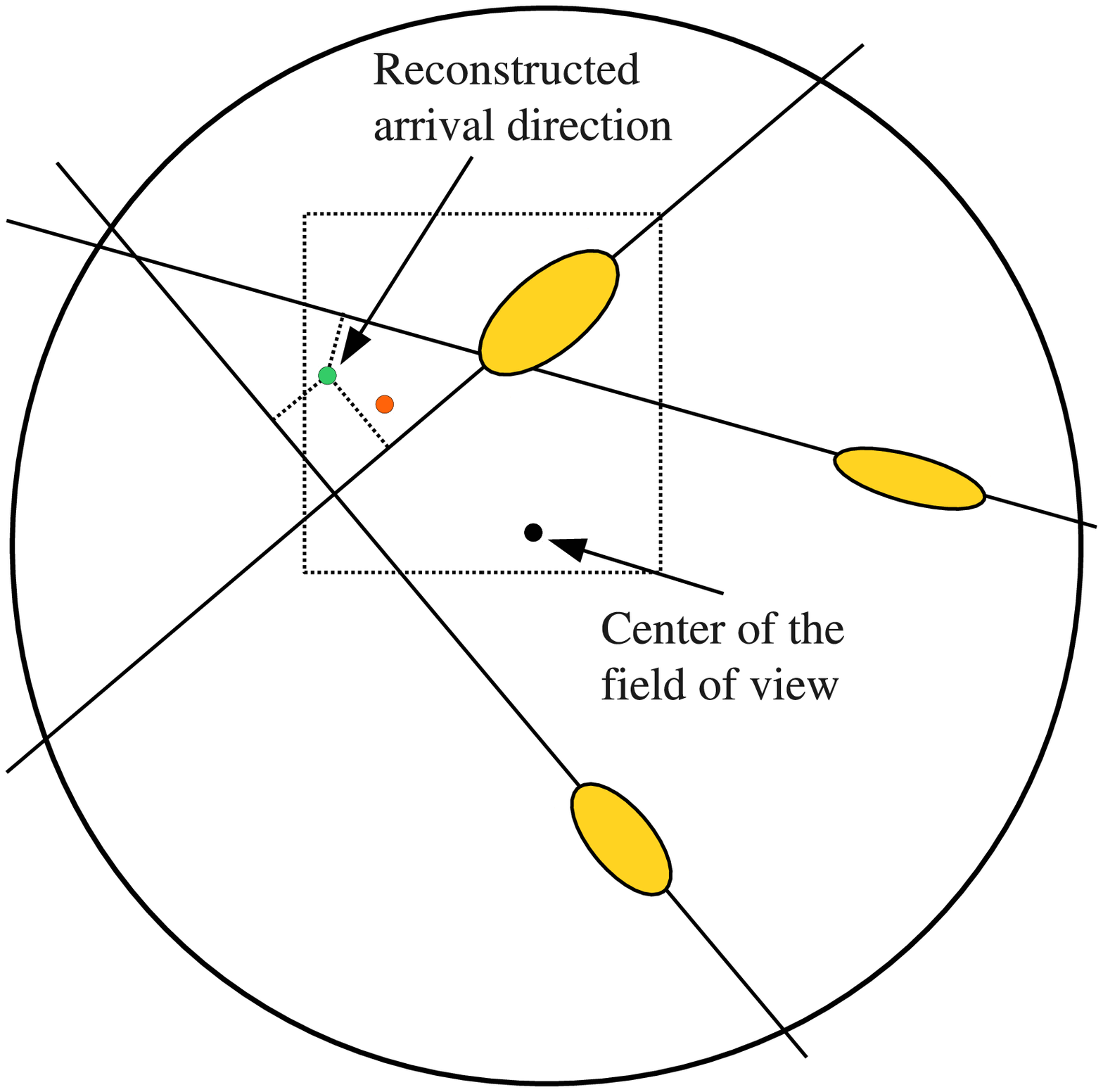}
  \includegraphics[width=2in]{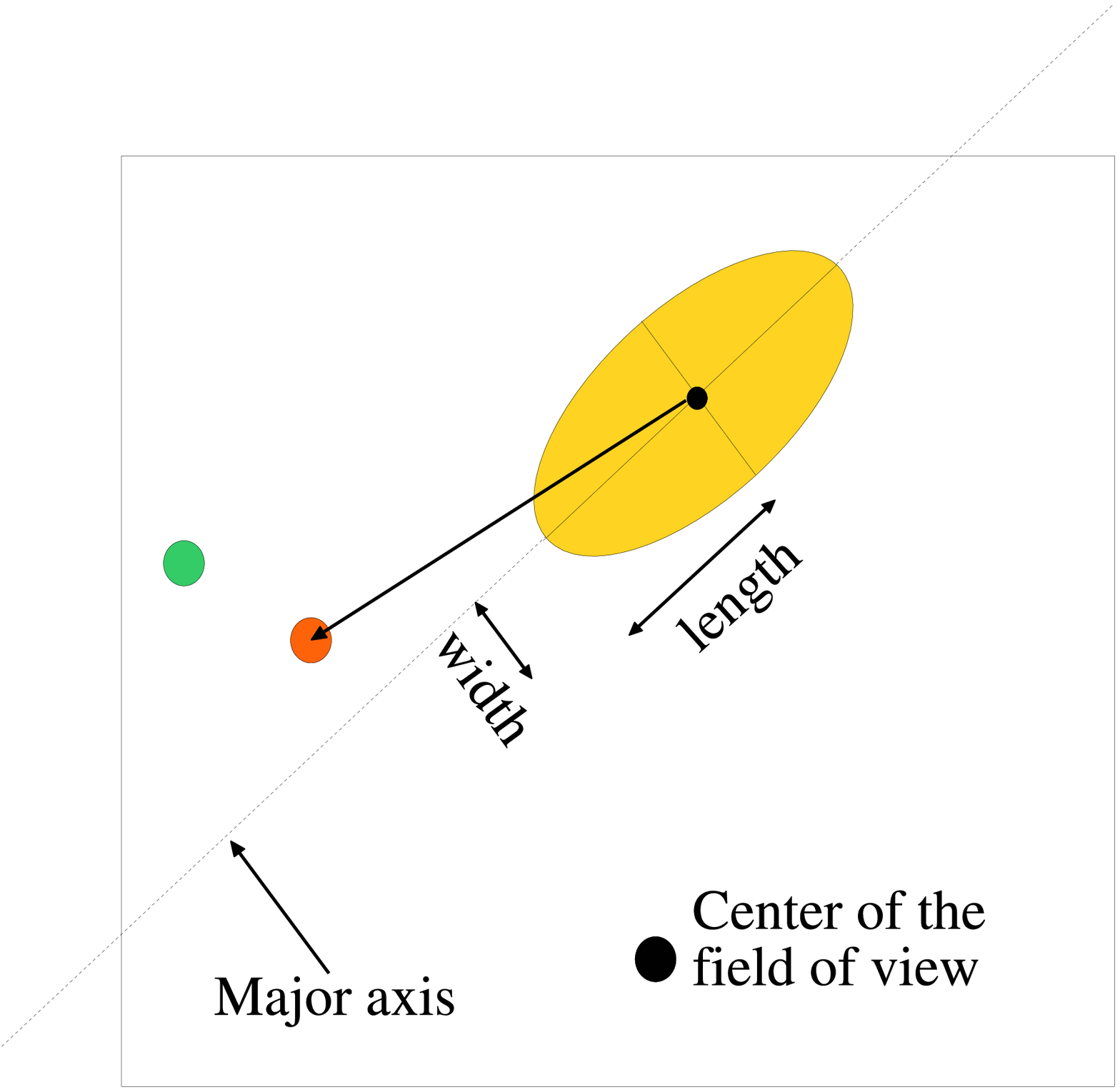}
  \includegraphics[width=2in]{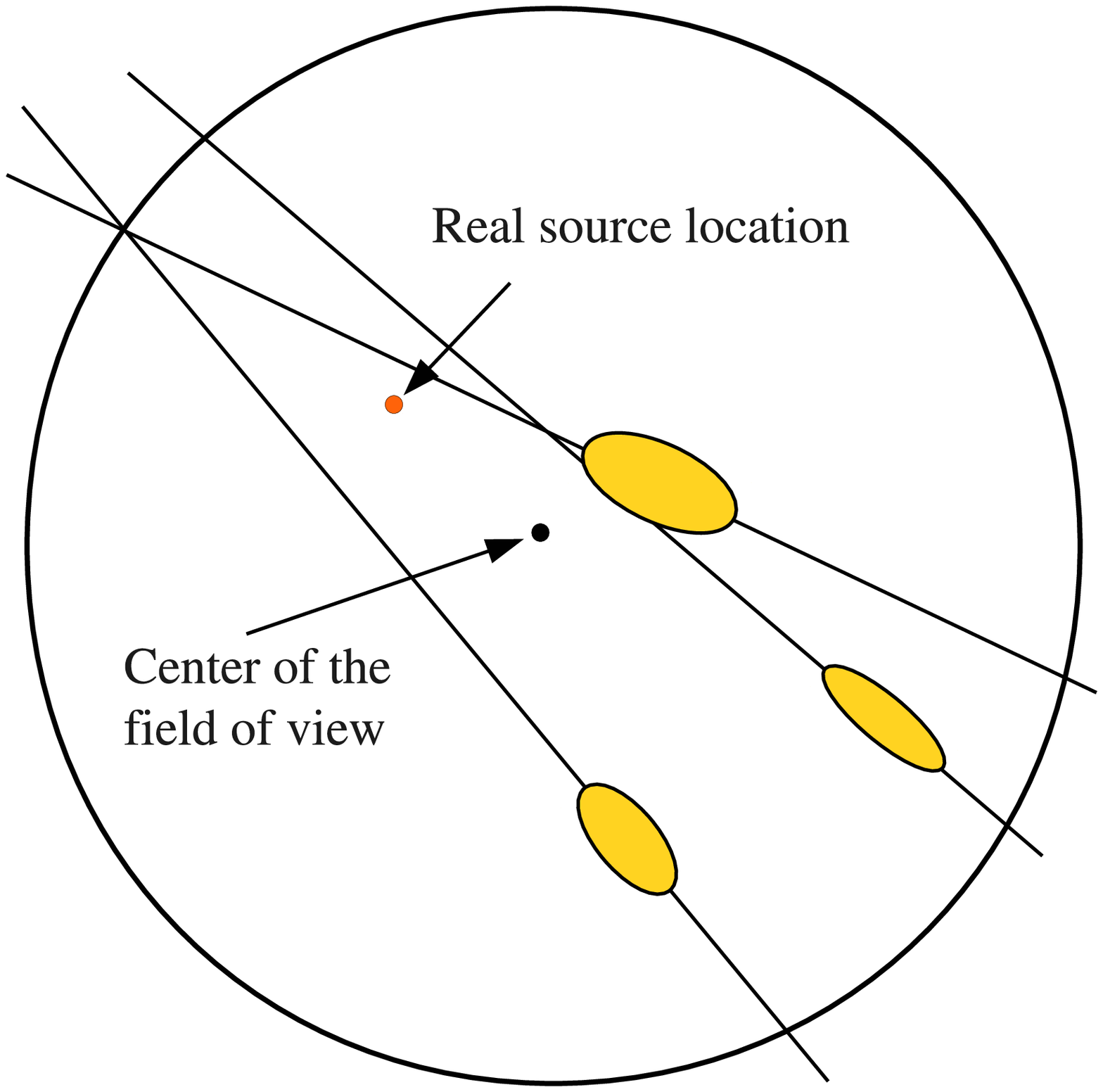}
  \caption{a) Representation of an air shower stereo image, seen by three different telescopes. b) Detail from a), $Hillas$ parameters that are used in the \emph{Disp} algorithm. c) Stereo images for LZA events tend to have mostly parallel major axes of ellipses. In all images, reconstructed and real source locations are represented by green and red dots respectively.}
  \label{wide_fig}
 \end{figure*}

\section{Description of the Algorithm}

The Disp method makes use of multidimensional lookup tables that contain the image information coming from Monte Carlo simulated air shower data. 
For each simulated shower image, the disp table stores the $Hillas$ parameters \emph{size}, \emph{width}, \emph{length} and \emph{disp}. 
These quantities are related to each other by the following geometrical argument: the projected images of air showers on the camera plane should look circular at the center, and become flatter as they move towards the edge. 
Therefore, \emph{disp} should increase with increasing ellipse flatness, which can be quantitatively described by \emph{width} and \emph{length} parameters.
The dependence on \emph{size} has been verified from parameter distribution plots.
Additional dimensions present in the lookup tables are zenith and azimuth angles, noise level and telescope ID.  
The way the reconstruction algorithm works is the following: for a given shower image, \emph{size}, \emph{width} and \emph{length} are calculated and the corresponding \emph{disp} is read from the lookup table. This tells us how far the arrival direction is from the image centroid along the major axis, but it does not tell at which side of the ellipse it is located, also known as head-tail ambiguity~\cite{hofmann99}.
To eliminate this problem, the closest cluster of points is picked, one coming from each ellipse.
In this way, the arrival direction is estimated for each telescope image and a weighted average is calculated.
The novelty of this implementation is that \emph{width} and \emph{length} parameters are being used as separate dimensions, as opposed to \emph{width}/\emph{length} parameter formerly used for this type of analysis~\cite{hofmann99}. 
Separating \emph{width} and \emph{length} yields better angular resolution for LZA data.

\section{Application \& Results for LZA Crab Nebula Observations}

The \emph{Disp} method has been tested on Monte Carlo simulations and LZA Crab data. 
A study on the zenith angle dependence of angular resolution (68\% containment radius) comparing the two reconstruction methods is shown in Figure 2. 
A significant improvement in the signal to noise ratio for data with zenith angles exceeding 50$^{\circ}$ is obtained. 
We measure an increase in significance of $\sim$20\% for a Crab-Nebula-like source and $\sim$30\% for a source having 1\% Crab Nebula strength. 
For small zenith angles, the geometrical method has the best angular resolution.

 \begin{figure}[th]
  \centering
  \includegraphics[width=3.in]{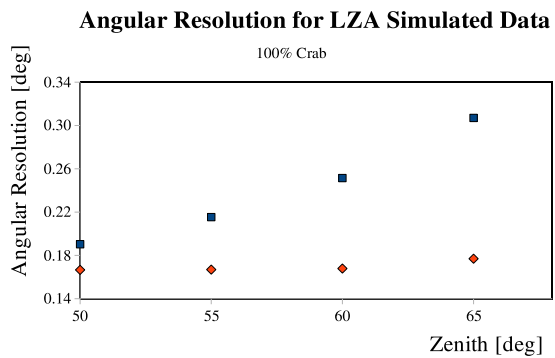}
  \caption{In the case of the default reconstruction algorithm, the angular resolution worsens with increasing zenith angle (blue squares). On the other hand, it remains mostly stable with the Disp method (red diamonds).}
  \label{wide_fig}
 \end{figure}

\section{Galactic Center Analysis}

VERITAS observed the Galactic Center at the position of Sgr A* for $\sim$15 hours with 62$^{\circ}$ average zenith angle. This is the first science application of the Disp method with VERITAS. The detection significance is more than 11~standard deviations \cite{beilicke2011}. Figure 3 shows the skymap for the VERITAS detection.

This research is supported by grants from the US Department of Energy, the US National Science Foundation, and the Smithsonian Institution, by NSERC in Canada, by Science Foundation Ireland (SFI 10/RFP/AST2748), and by STFC in the UK. 
We acknowledge the excellent work of the technical support staff at the FLWO and the collaborating institutions in the construction and
operation of the instrument.

 \begin{figure}[!t]
  \vspace{5mm}
  \centering
  \includegraphics[width=2.in]{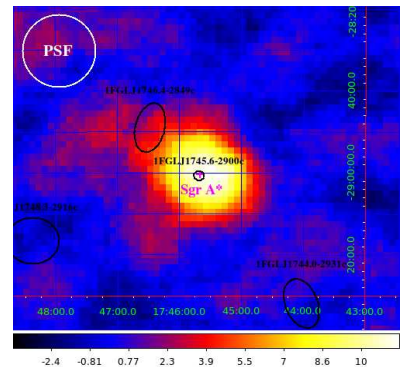}
  \caption{VERITAS significance map for Sgr A*. \emph{Fermi}-LAT ellipses are shown in black. The 68\% containment radius is 0.13$^{\circ}$ (White circle at the top left corner).}
  \label{simp_fig}
 \end{figure}


\clearpage

\end{document}